
\input phyzzx
\input table
\def\ts{\thinspace}
\def\refmark#1{{[#1]}}
\REF\INF{K.~Sato \journal Phys. Lett.&99B(81) 66
  \journal Month. Not. R. Astron. Soc.&195(81) 467. \subpar
 A.~H.~Guth \journal Phys. Rev.&D23 (81)347.}
\REF\M{K.~Maeda, K.~Sato, M.~Sasaki and H.~Kodama \journal Phys. Lett.&108B(82)
 98.}
\REF\NS{Y.~Nambu and M.~Siino \journal Phys. Rev.&D46(92) 5367.}
\REF\H{S.~W.~Hawking \journal Commun. Math. Phys. &43(75)199.}
\REF\VA{P.~C.~Vaidya \journal Proc. Indian Acad. Sci. &A33(51)264}
\REF\W{See, e.g., R.~M. Wald, \ts {\sl~General Relativity} (\ts University of
Chicago
 Press, Chicago, 1984).}
\REF\PG{D.~S.~Goldwirth and T.~Piran \journal Phys. Rev.&D40(89) 3263.}
\REF\HE{See, e.g., S.~W.~Hawking and G.~F.~R.~Ellis, \ts {\sl~The large
scale structure of space-timennnnn} (\ts Cambridge University
 Press, Cambridge, 1973).}
\REF\G{S.~K.~Blau, E.~I.~Guendelman and A.~H.~Guth \journal Phys. Rev.&D35(87)
 1747.}
\REF\MASS{W. Fishler, D. Morgan and J. Polchinski \journal Phys.
Rev.&D42(90)4042.}
\REF\CO{W.~Collins \journal Phys. Rev.&D45(92) 495.}
\FIG\a{
(a) When a Schwarzschild mass decreases, the AH becomes timelike. (b) With
an increasing Schwarzschild mass, the AH is spacelike.}
\FIG\b{
(a) In the de Sitter universe the ingoing cosmological AH becomes an ingoing
null surface. (b) There is the ingoing cosmological AH directed to
timelike direction in the dust universe. (c) The ingoing cosmological
AH coincides with an outgoing null surface.}
\FIG\c{
The time evolution of the scalar field.}
\FIG\d{Time evolution of a worm hole.
`$\cdot$'\ belongs to the region R1\ts ($\theta_+>0, \theta_-<0$),
`$\times$'\  to the region R2\ts ($\theta_+<0,\theta_-<0$),
`$\bullet$'\ to the region R3\ts ($\theta_+>0, \theta_->0$) and `$\circ$'\
to the region R4\ts ($\theta_+<0, \theta_->0$). When $t=120$, a Schwarzschild
AH exists.}
\FIG\e{
The worm hole spacetime has three AHs. A Schwarzschild AH is a boundary
between the region R2 `$\times$' and the region R4 `$\circ$'. An ingoing
cosmological AH is a boundary between the region R1 `$\cdot$' and the
region R3 `$\bullet$'. An outgoing
cosmological AH is a boundary between the region R3 `$\bullet$' and the
region R4 `$\circ$'.}
\FIG\f{An AH becomes ingoing null surface and coincides with de Sitter
event horizon.}
\FIG\g{From $t_0$ to $t_1$, a Schwarzschild AH is spacelike. After $t_1$ it
becomes close to null. At $t=t_2$, the Schwarzschild AH changes its
direction to timelike.}
\FIG\h{
The distribution of the local mass and the location of the AH at
(a) t=0$\sim$10.0,
(b) t=80.0$\sim$90.0,
(c) t=115.0$\sim$125.0 and
(d) t=140.0$\sim$150.0.
}
\FIG\i{
Dots are plotted at grids where the outgoing energy flux is large. The
trajectory of the dots crosses Schwarzschild AH at
$t=t_2$.}
\FIG\j{(a) The ingoing cosmological AH. (b) The outgoing
cosmological AH. Both graphs tell that AHs are timelike.}
\FIG\k{The changes of the $p/\rho$ at $\chi=1, 5, 9, 20, 200, 393$.
The matter field becomes dust like at $\chi=1, 5$. At $\chi=9, 20, 200, 393$,
 the contribution of the energy flux exists.}
%
%
%
\pubnum={DPNU-93-    \cr
        TIT/HEP232-COSMO36}
\date={August 1993}
\titlepage
\title{Local Structure of Numerically Generated Worm Hole Spacetime}
\vfill
\author{Yasusada NAMBU\footnote{{}^\dagger}
{E-mail address: nambu@jpnyitp.bitnet} and Masaru SIINO\footnote{{}^\ddagger}
{JSPS fellow, E-mail address: msiino@phys.titech.ac.jp}}
\address{${}^\dagger$Department of Physics, Nagoya University \break
Chikusaku Nagoya 464-01,
Japan}
\address{${}^\ddagger$Department of Physics, Tokyo Institute of
Technology \break
 Oh-Okayama, Meguroku, Tokyo 152, Japan}
\vfill
\abstract{
We investigate the evolution of the apparent horizons in a numerically
gererated worm hole spacetime. The behavior of the apparent horizons
is affected by the dynamics of the matter field. By using the
local mass of the system, we interpret the evolution of the worm hole
structure.
}
\endpage
%
\hsize=15.1cm
\chapter{Introduction}

The inflationary scenario succeeded in explaining the origin of global
homogeneity\refmark\INF .
 An rapid cosmic expansion erases out initial inhomogeneity and provides
 the present homogeneous universe. However, it may produce some new
inhomogeneity whether they are desirable or not.
It is expected that the inflationary expansion makes density perturbation
from a quantum fluctuation; this is considered as the seed of the galaxy
observed as present local inhomogeneity.
Inflation also has a possibility of creating a global inhomogeneous structure
 like a worm hole.

  In a thin wall approximation and the context of the old inflation,
  Maeda, Sato, Sasaki and Kodama\refmark\M ~demonstrated that the worm
  hole is formed in the spherically symmetric spacetime. By their work one can
decide whether a given initial universe evolves to a worm hole
space-time.  In our previous paper\refmark\NS ~these aspects were confirmed
 numerically  for a closed universe with a scalar field.
 There is no remarkable difference between the thin wall analysis and our
numerical calculation: both these works insist that a worm hole
formation occurs when  suitable initial conditions are satisfied.
But there exists qualitative  differences between  these
two models.
 In the worm hole with the thin wall approximation, a Schwarzschild
 mass is a constant parameter because the
local  structure of the spacetime
 is static. There is no local dynamics, and only the global dynamics
 of a worm hole exists. On the contrary, there is local
 dynamics in the numerically generated worm hole because it contains
 the scalar field.
 The evolution of our numerical worm hole is strongly affected by the
 dynamics of the scalar field.
 In our previous work, we determined the apparent
 horizons (AHs) for outgoing and ingoing null geodesics to judge
  a worm hole formation occurs. Since these AHs
 depend only on local geometry, we can relate the behavior of AH to
 the local dynamics.
 The Schwarzschild AH which indicates the existence of the
worm hole is related to the evolution of the Schwarzschild mass.
 The cosmological AHs appearing in the locally expanding region
 are related to an expanding geometry.

 In this paper we concentrate on the dynamics of AHs in a worm hole
 space time. We calculate null geodesics
 around each AH to decide the causal direction of the AH.
 The causal direction of the AH is compared with that
of some typical space-times. The results will give us information about
 the local  dynamics. Our purpose is to relate the locally dynamics of
 the matter and the evolution of the geometry.

 This paper is organized as follows. Section 2 gives some examples of the
 AHs in the dynamical spacetime; they are compared with our
 numerical spacetime. In section 3,  the worm hole spacetime is numerically
 generated. We investigate the behavior of the AHs in
 section 4. The final section is devoted to summary and discussion.

\chapter{Apparent Horizon in the Dynamical Spacetime}

In this section, we consider space-times whose metrics are
analytically known. We calculate the motion of AHs and
investigate what determines the causal
direction(timelike, null, spacelike)  of AHs trajectory.

  First, we give the Vaidya\refmark\VA ~spacetime as an example of a black
 hole solution whose
  Schwarzschild mass $M$ depends on outgoing null
  coordinate $v$. The metric is given by
 $$
  ds^2=-\Biggl(1-{2 M(v) \over r}\Biggr) dv^2 + 2 dv dr + r^2 d \Omega^2 \ \ \
.\eqno\eq
  $$
 In this spacetime an AH for outgoing null geodesic is expressed as
  $$
  r=2 M(v) \ \  .\eqno\eq
  $$
  Fig.\a (a) (Fig.\a (b)) shows a Penrose diagram for the case that
Schwarzschild mass $M$\ is decreasing (increasing) function of $v$.
  If the mass decreases (increases), the apparent
  horizon is directed to a timelike (spacelike). The decreasing
  Schwarzschild mass implies the violation of the energy condition.
  Of course, when Schwarzschild mass is constant, the spacetime is
  static and the AH becomes an outgoing null surface.

 Second, we consider cosmological AHs in a
  closed universe(we consider only closed universe because of the
  technical reasons \refmark\NS). They are determined for both
outgoing null geodesics and
  ingoing null geodesics. Nevertheless, we show the
  cosmological AH for only ingoing null geodesic here because ``ingoing'' and
``outgoing'' are symmetric for homogeneous and isotropic cosmological model.
  Metrics of closed Robertson-Walker universe is
  $$
  ds^2=-dt^2 + R(t)^2 ( d\chi^2 + \sin^2 \chi d\Omega^2 ) \ \ \ ,\eqno\eq
  $$
  and its AH is given by the equation
  $$
  \tan \chi= \pm {1 \over \dot R(t)} \ \ \ ,\eqno\eq
  $$
where the signature $+$($-$) is selected for the ingoing (outgoing)
apparent  horizon.
  The Penrose diagrams of the universe dominated by vacuum energy,
  dust and radiation are shown in Fig.\b (a)-(c),
 respectively.
  When the vacuum energy dominates,
  the spacetime is de Sitter and an ingoing AH becomes an ingoing null
  surface. Fig.\b (b) tells us that the AH has a timelike direction in the
 dust universe. For the universe with radiation, the ingoing AH coincides with
  an outgoing null surface.

\chapter{Numerically Generated Worm Hole Spacetime}
In our previous work\refmark\NS, we showed that a worm hole formations is
possible for suitable initial conditions.
Here we consider one typical case of the worm hole formations.

 The metric used by our calculation is
$$
 \eqalign{
  ds^2 = -dt^2 + R(\chi,t)^2 \Bigl\lbrack (d\chi+\beta dt)^2 +
     \sin^2\chi d\Omega^2 \Bigr\rbrack      ,\cr
         0\leq \chi \leq \pi \ , }\eqn\METST
$$
where  $R(\chi,t)$\ is an inhomogeneous scale factor that is a simple
generalization of the scale factor $R(t)$ of the FRW metric and $\beta$ is
a shift function. Using this metric form, we solve the full Einstein equation
 by finite differencing on numerical grid\refmark\PG~. We see only a half of
 spacetime $\chi=0\sim\pi/2$ by imposing a reflection symmetry at $\chi=\pi/2$
{}.

 As the typical case of the worm hole formations, we have used the
following form of the initial distribution for the massive scalar
field(Fig.\c):
$$
  \Phi = \Phi_0 + {A\over 1-e^{-{1\over D^2}}}\bigl[
  \exp (-({\cos\chi\over D})^2)- \exp (-{1\over D^2}) \bigr].
   \eqno\eq
$$
In this model, the universe contains one false vacuum ``bubble''\refmark\G
{}~(``bubble'' means the region that is nearly homogeneous around
$\chi=0$)
surrounded by a true vacuum region. The potential of the massive scalar field
  $\Phi$ is $V(\Phi)=\lambda (\Phi^2-\sigma^2)^2$ with $\sigma=1, \lambda=1$.
Parameters are $A=0.35, D=0.2, \Phi_0=0.045$\ and $ K_0=-2.4$.
A spatial grid size and a time interval are set to $\pi/800$ and $0.005$,
respectively. The relative error in the Hamiltonian constraint is below
$5\times 10^{-3}$\  during the run\refmark\NS.

In this calculation, we see that a worm hole is formed.
The time evolution of the scalar field is shown in Fig.\c.
 In Fig.\d ~we embed each time slices into a higher dimensional flat space.
 Every grid points ($\chi=0\sim\pi/2$)\ are plotted by solving the embedding
 equation.
 At the beginning of the calculation, the universe is a homogeneous 3-sphere.
 As the universe evolves, the shape of the time slice is deformed and yields a
 bottle necked structure. To exhibit the information about the AH,
  we draw each grid point on a time slice by four kinds of symbols
  which are distinguished by the   sign of the expansions  $\theta_+$
  and $\theta_-$:
\vskip 0.5 cm
\begintable
 region name | sign of expansion      | symbol \cr
   R1        |$\theta_+>0, \theta_-<0$ |\hfil$\cdot$ \hfil\cr
   R2        |$\theta_+<0, \theta_-<0$ |\hfil$\times$\hfil\cr
   R3        |$\theta_+>0, \theta_->0$ |\hfil$\bullet$\hfil\cr
   R4        |$\theta_+<0, \theta_->0$ |\hfil$\circ$\hfil
\endtable
\noindent%
R3($\theta_\pm >0$) is the region between the ingoing
cosmological AH and the outgoing cosmological AH.
 R2($\theta_\pm <0$) is the inner region of the Schwarzschild apparent
horizon\ts (inside the black hole horizon). At the early stage of the
time steps, we can see the de Sitter AH which is located at the
boundary  between the region  R1 and the region R3. When $t=120$,
 a Schwarzschild AH appears at the boundary between the
 region R2 and R4. At this time, we can say that an Einstein-Rosen bridge is
formed. It connects causally disconnected two cosmological bubble regions: this
 structure is a worm hole.

 Fig.\e ~is the trajectories of the AHs on a
$t$-$\chi$ plane. The three boundary lines separating each region are the
trajectories of the AHs. Their behavior is investigated  in the
next section.
\chapter{Null Geodesic and Apparent horizons}

An outgoing (ingoing) null vector field is given by
$$
k^a=\Bigl({\partial \over \partial t}\Bigr)^a+\bigl(-\beta \pm{1 \over
R}\bigr)
\Bigl({\partial \over \partial \chi}\Bigr)^a \ \ \ .\eqno\eq
  $$
All outgoing (ingoing) null geodesics are generated by the exponential
map\refmark\HE~ of this null
vector field. Then the trajectory of the null geodesics becomes
$$
\chi_{null}(t)-\chi_{null}(t_i)=\int_{t_i}^{t}(-\beta(\chi_{null}(t^\prime),t^\prime)\pm
 {1 \over R(\chi_{null}(t^\prime),t^\prime)}) dt^\prime \ \ \ .\eqno\eq
 $$
For a check of our numerical analysis,we first calculate the ingoing
null geodesics in a numerically generated
de Sitter spacetime(Fig.6).
 There are two types of null geodesics. One type can reach $\chi=0$
 and the other cannot reach $\chi=0$. The ingoing AH coincides with
 the  boundary of these two type of geodesics. This agrees with the
 fact that an AH becomes ingoing null surface in de Sitter spacetime.
 We find this ingoing null geodesic generates event horizon.

 In our numerical worm hole spacetime ($0\leq\chi\leq \pi/2$),
 there are three AHs: an ingoing cosmological AH,
an outgoing cosmological AH and an ingoing Schwarzschild
 AH (see Fig.\e). The ``ingoing (outgoing)'' means a direction in which
 $\chi$ decreases (increases). Fig.\g ~shows null geodesics around the
 Schwarzschild apparent horizon.
 When the Schwarzschild AH appears at $t= t_0$, it moves
 to an ingoing space-like direction.
 After  $t= t_2$,
 its direction becomes timelike. Comparing these
 aspects with the Vaidya spacetimes (see Fig.\a), we expect that
 the ``Schwarzschild'' mass of the  worm hole increases and AH becomes space
like initially.
After $t=t_2$, the mass decreases and AH becomes time like.

To see the behavior of AH more precisely, we introduce the local mass.
In spherically symmetric space-time, we can define a local mass\refmark\MASS:
 $$
 {\cal M}(\chi)={1 \over 8R\sin\chi}(\theta_+\theta_-+4R^2\sin^2\chi) \ \ \ .
\eqn\LM
 $$
This mass function coincides with the usual ADM mass in the asymptotically
flat space-time.  The behavior of the Schwarzschild AH is
decided by this local mass because it reflects the dynamics of the
local matter field.
 Fig.\h ~shows the evolution of ${\cal M}(\chi)$. At the beginning of
 the evolution (Fig.\h (a)), the
 distribution of ${\cal M}(\chi)$ quickly approaches to a typical form.
 The peak of ${\cal M}(\chi)$ exists on the position of the domain
 wall which separate the false and the true vacuum regions.
The value of ${\cal M}(\chi)$ in the bottle necked region is important.
 In Fig.\h(a), ${\cal M}(\chi\sim\pi /2)$ settles down to a finite value and
 slowly increases. When the Schwarzschild AH is formed, ${\cal M}(\chi)$
 evolves as Fig.\h(b). The peak is shifting to left because we do not
 choose a comoving gauge.
We observe that ${\cal M}(\chi)$ becomes negative at the right-hand
side of the peak.
 Fig.\h(c) shows that the
 value of ${\cal M}(\chi)$ at the center of the bottle necked structure
 suddenly begin to decrease. In the final stage of the evolution
 (Fig.\h(d)), the distribution of ${\cal M}(\chi)$ does not change so much
 at the bottle necked structure but the value ${\cal M}(\chi)$ continuously
decreases.

 From the Einstein equation, we get a relation between ${\cal M}(\chi)$ and the
 energy-momentum tensor of the matter field;
 $$
 {\cal M}(\chi)=\int_{ }^\chi\Biggl ({(R\sin\chi),_\chi \over R} {\cal
   H}_{tM} -
 {\dot{R}\sin\chi - \beta (R\sin\chi),_\chi \over R} {\cal
   H}_{rM}\Biggr) d\chi\ \ \ ,      \eqn\IM
 $$
 where ${\cal H}_{tM}$ and ${\cal H}_{rM}$ are the
 contribution of the matter field in the  Hamiltonian constraint(energy
density) and the momentum constraint(energy flux), respectively. The first
term of \IM\ can be rewritten to
 $$
 \eqalign{
  & \int_0^\chi {(R\sin\chi),_\chi \over R} {\cal H}_{tM}d\chi \cr
 &=\int_0^{R\sin\chi\vert_{\chi=\chi_{max}
}}\rho(\chi)(R\sin\chi)^2d(R\sin\chi)-\int_{R\sin\chi\vert_{\chi=\chi_{max}
}}^{R\sin\chi}\rho(\chi)(R\sin\chi)^2d(R\sin\chi)\ \ \  ,}   \eqn\RM
 $$
where $\chi=\chi_{max}$ gives the maximum value of $R(\chi)\sin\chi$.
In Fig.\d, as the horizontal axis is
$R\sin\chi$, we can see that
$\partial(R\sin\chi) / \partial\chi$ is positive for $\chi<\chi_{max}$
and is negative for $\chi> \chi_{max}$. Therefore the first
term in Eq.\IM\ is positive, by two inequalities,
$\rho(\chi<\chi_{max})>\rho(\chi>\chi_{max})$ and $R\sin\chi>0$.
 The negative ${\cal M}(\chi)$ implies that an outgoing energy flux
(the second term in local mass)  dominates ${\cal M}(\chi)$ there.
 This is checked by observing the locations where the outgoing energy
 flux is large on
 ($t-\chi$) plane at each time-slices.
This energy flux is mainly carried by the domain wall.

By using the local mass ${\cal M}(\chi)$ and the energy flux we can
 explain the change of the causal direction of the Schwarzschild AH. When
 the Schwarzschild AH appears, it is spacelike. At this time, the value of
${\cal M}(\chi)$  in the bottle necked region increases.
  Fig.\i ~tells that the energy flux is not
 important in this region and  the increase
of ${\cal M}({\rm R2})$ is
 caused by the evolution of the energy density $\rho$; at this time, the
 $\rho$ of the bottle necked region decreases more
 quickly than that of
 the cosmological bubble region. When the Schwarzschild AH changes its
 causal direction from spacelike to
 timelike, ${\cal M}(\chi =\pi/2)$ begin to decrease.
 This is caused by the energy flux of the matter field.
 After the energy flux crosses the
 Schwarzschild AH, the energy density in the region R2($\theta_\pm<0$)
 increases.
This implies that  the contribution of the
first term in \RM ~becomes smaller and that of the second term in \RM ~becomes
larger. But the energy flux is not so large as to
dominate the evolution
of ${\cal M}(\chi)$ (see Fig.\i). Hence, ${\cal M}({\rm R2})$ becomes small.
 This flow continues till
 the end of the calculation. These facts agree with the guess from the
 Vaidya spacetime if we especially regard ${\cal M}({\rm on the
Schwarzschild AH})$ as the
 Vaidya's Schwarzschild mass. As the radius of the worm hole is given by $r=
2G{\cal M}$, the decreasing mass on the Schwarzschild AH means that the worm
hole is
shrinking.

 In Fig.\j(a) (Fig.\j(b)), ingoing and outgoing null geodesics are
 drawn around the
 ingoing (outgoing) cosmological AH. In the both figures, the
 cosmological AHs are
 timelike at the final stage of the calculation. Reminding the result
 of section 2 on cosmological AH, we can say that
 the local spacetime around these cosmological AHs is similar to the
 dust universe.
 This is confirmed by investigating the time dependence of ratios
 $p/\rho$ at some spatial grid points (see Fig.\k).
 Since the scalar field begins to oscillate, we average out this value for a
 few cycles.
 The equation of state of the matter settle dawn to
 that of the dust ($p=0$) at the grid points $\chi/\delta \chi=1, 5$. At
 the other grid points
 $\chi/\delta\chi=9, 20, 200, 393$, the contribution of the energy flux is not
 negligible and the local geometry is not dust universe like.

\chapter{Summary and Discussion}

We have investigated the AHs of the numerically
generated worm hole spacetime. This tells us about both
the global structure and the local structure of the spacetime.
In previous paper\refmark\NS, we discussed about this global structure
and judged  whether a worm hole structure is formed.
In this paper, we concentrate on the local structure and determine
the causal direction of the AHs.

 We have three AHs in a half of space  $0\leq\chi\leq\pi/2$;
one of them is Schwarzschild one and two of them are cosmological ones.
The  trajectory of the Schwarzschild AH is  spacelike
 when it appears. The size of the worm hole grows because the area
 of AH increases.  After then, its  direction becomes close to null.
 At $t=t_2$ it becomes timelike.
 At this time, the domain wall crosses the Schwarzschild AH, and the
 Schwarzschild  mass of the Einstein-Rosen bridge begins to decrease.
Considering the local mass on the Schwarzschild AH as the substitute of the
Vaidya's Schwarzschild
 mass, its behavior agrees with the aspects of the evaporating Vaidya
space-time. By observing the local mass ${\cal M}(\chi)$ and the energy flux of
 the scalar field, we can confirm this interpretations.
 The behavior of the cosmological AHs depends on local condition
 of the matter field. Comparing with some typical cosmological model
 (Fig.\b), the cosmological bubble region
 becomes dust like universe at the final stage of the numerical simulation.
 This is explained by the fact that the scalar
 field settles down to a coherently oscillating state in this region.

In this paper, we observed that the causal direction of AH is related to
increase or decrease of the local mass. AH becomes spacelike when the mass
increases and becomes timelike when the mass decreases. It is possible to
 prove this feature of AH for a spherical spacetime\refmark\CO.

\ack

 We would like to thank Prof. A.~Hosoya for his continuous encouragement and
 K. Nakamura for discussing about local mass.
The calculations have been performed by SUN SPARK2.
 One of the authors (M.S.) thanks the Japan Society for the
 Promotion of Science for financial support. This work was supported in part
 by the Japanese Grant-in-Aid for Scientific Research Fund of the Ministry of
 Education, Science and Culture.
\endpage

\refout
\figout
\bye